\documentstyle[aas2pp4]{article}
\newcommand{\grad}{\mathop{\rm grad}\nolimits}

\begin{document}

\title{
Three-dimensional Modeling of the Matter Flow Structure in Semidetached
Binary Systems}

\author{D.V.Bisikalo\altaffilmark{1}}
\affil{Institute of Astronomy of the Russian Acad. of Sci.,
Moscow, Russia}

\author{A.A.Boyarchuk}
\affil{Institute of Astronomy of the Russian Acad. of Sci.,
Moscow, Russia}

\author{O.A.Kuznetsov\altaffilmark{2}}
\affil{Keldysh Institute of Applied Mathematics, Moscow, Russia}

\author{V.M.Chechetkin}
\affil{Keldysh Institute of Applied Mathematics, Moscow, Russia}

\altaffiltext{1}{\large E-mail address: {\it bisikalo@inasan.rssi.ru}}
\altaffiltext{2}{\large E-mail address: {\it kuznecov@spp.keldysh.ru}}

\begin{abstract}

Results of numerical simulations of matter flows in a
semidetached binary system similar to the low-mass X-ray binary
X1822--371 are presented. Three-dimensional modeling of the mass
transfer gas dynamics makes it possible to investigate gas
streams in the system and to study the influence of a common
envelope. The presence of the common envelope leads to the
absence of shock interaction between the stream of matter
flowing from the inner Lagrange point and the gas in the
accretion disk. The stream is deflected by the gas in the common
envelope and approaches the disk tangentially, so that it does
not cause any shock perturbation ("hot spot") on the disk. At
the same time, the interaction of the stream with the common
envelope leads to the formation of an extended shock wave along
the edge of the stream. The observational manifestation of this
shock is estimated to be equivalent to that of a hot spot in the
disk.  The calculated accretion disk parameters are presented.

\end{abstract}

\section{Introduction}

   Semidetached binary systems are interacting binaries, in
which one of the stars fills its critical surface, resulting in
mass exchange between the components of the system. In general,
the critical surface has a complicated shape [1], and special
mathematical models are required to describe matter flows in the
system (a review of such models is given in [2]).  However, the
large amount of observational data confirming the small
eccentricity of the orbits and the high degree of
synchronization in these binary systems makes it possible to
simplify the models used. In the standard formulation, the
components of semidetached binary systems are assumed to have
circular orbits and to rotate in synchrony with the orbital
motion. In this case, the critical surface is identified with
the inner surface (Roche surface) in the restricted three-body
problem, and it is assumed that the mass exchange between the
system components occurs through the vicinity of the inner
Lagrange point $L_1$, where the pressure gradient is not
balanced by the force of gravity.

   The gas dynamics of mass transfer through the inner Lagrange
point $L_1$ have been studied by many authors. A thorough
analysis of the flow of matter in the vicinity of $L_1$ was
performed by Lubow and Shu [3], who estimated the basic
characteristics of the flow in a semi-analytical approximation
using a perturbation method. The small parameter $\epsilon$
used in this analysis was the ratio of the sound speed $c_0$ at
the star's surface to the characteristic orbital velocity
$\Omega A$ ($\Omega$ is the angular velocity of rotation of the
system components and $A$ is the distance between the component
centers). In another approach based on Bernoulli integral
analysis, the parameters of the stream of matter were refined,
and a relationship obtained between the mass transfer rate and
the degree of overfilling of the Roche lobe by the donor star
[4,5].

  To adequately describe the mass exchange process, it is
necessary, in addition to determining the stream parameters, to
consider the further behavior of the stream lines after the
motion of matter from $L_1$. It is the mass transfer process that
forms the overall pattern of the flow and, consequently, defines
its basic observational manifestations; therefore, much
attention has been given to the study of this problem. Warner
and Peters [6], Lubow and Shu [3], and Flannery [7] pioneered
the investigation of particles ejected from $L_1$ and moving in
the gravitation field of the binary system. The results obtained
in these and a number of subsequent studies have gained wide
recognition, but their application to the analyses of specific
systems led to the discovery of substantial discrepancies with
the observational data.  These discrepancies are associated with
the use of a simplified ballistic approach in the gas motion
analysis, which did not take into account gas dynamics effects.
To consider the influence on gas motions of a circumstellar
envelope during and after its formation, and, consequently, to
accurately describe matter flows, it is necessary to solve a
complete system of gas dynamics equations; this is possible only
using quite sophisticated mathematical models.

   The use of numerical methods to study the gas dynamics of
mass transfer in semidetached binary system was for a long time
limited by the development of computer techniques, so that
two-dimensional (2D) models were used for the flow analysis. In
spite of the simplification of the 2D approach, it enabled a
correct treatment of some particular features of the matter
flow, and made it possible to obtain a number of interesting
results (see, e.g., [8--12]). In recent years, numerical
investigations of mass transfer gas dynamics using more
realistic three-dimensional (3D) models have become possible
[13--20]. Such studies have considered, in particular, accretion
disk formation in semidetached binary systems [13, 15], and also the
region of interaction between the stream flowing from $L_1$ and
the disk [14, 20]. Unfortunately, these studies were conducted
in a limited formulation, without considering the process of
establishing the flow on long time scales; this prevented
investigation of the real morphology of the systems and,
correspondingly, evaluation of the impact of a common envelope
on the flow pattern. Some progress in the investigation of the
overall flow pattern was achieved in the studies of Molteni,
Belvedere, and Lanzafame [16--19], who conducted 3D numerical
simulations over rather long time intervals. A number of
interesting results were obtained in these papers, however, the
use of the Smoothed Particle Hydrodynamics (SPH) method to solve
the system of gas dynamics equations made it impossible to study
the influence of a common envelope on the flow pattern. The
restrictions imposed by the SPH method do not allow
investigation of flows with large density gradients, and,
therefore, the action of the rarefied envelope gas on the mass
transfer gas dynamics was not taken into account entirely
correctly.

   Here, we present the results of 3D numerical simulations of
the flow patterns in semidetached binary systems. The Total
Variation Diminishing (TVD) method we use to solve the system of
gas dynamics equations does not have the disadvantages of the
SPH method. This made it possible to investigate gas flow
morphology in the system and the effect of the common envelope
of the system, despite the presence of substantial density
gradients. We studied the gas dynamics of matter flows in
semidetached binary systems over long time intervals, which
enabled us to consider the basic features of the flow pattern in
a steady-state (stationary) regime.

\section{The Model}

\subsection{Physical Model}

   Semidetached binary systems exhibiting extremely interesting
observational manifestations of their interaction include
cataclysmic binaries, low mass X-ray binaries (LMXB), and
supersoft X-ray sources. The available observations of LMXB
X-ray light curves provide striking evidence for the presence of
complicated flow patterns, and suggest some characteristic
features of the gas flow structure. In particular, in a number
of X-ray "dipping" sources, of which the most widely known is
the X1822--371 system, a substantial decrease of the emission is
observed at phase 0.8 of the orbital period, associated with a
thickening in the disk in the place where it is perturbed by the
stream of gas flowing from $L_1$ [20--22].

   In our studies of matter flow patterns,  we considered
semidetached binary systems with parameters typical for low mass
X-ray binaries, where a main sequence dwarf fills its Roche lobe
and transfers mass to a neutron star. For more specific
calculations, we chose a system with parameters close to those
of X1822--371 [20]. We assumed that the primary star filling its Roche lobe
has a mass $M_1$ equal to $0.28 M_\odot$ and a surface gas
temperature $T = 10^4 K$; the mass of the companion (a compact
object with radius $0.05 R_\odot$) has $M_2=1.4 M_\odot$; the
orbital period of the system is $P_{orb} = 1^d78$; and the
distance between the centers of the two stars is $A =
7.35R_\odot$.

    To adequately describe the matter flow structure in the
binary system we have chosen, we must take into account the
effect of radiation processes upon the gas dynamics. Including
non-adiabatic processes in the numerical model substantially
increases the calculation time. For this reason, based on the
available computational resources and the need to let the
calculations run over long time intervals (to obtain a
steady-state regime for the flow), we restricted our
consideration to an ideal gas model with adiabatic index
$\gamma$ and equation of state $P = (\gamma-1)\rho\varepsilon$
(where $P$ is pressure, $\rho$ is density, and $\varepsilon$ is
specific internal energy).  The ideal gas assumption in the
model formally neglects radiative losses; however, in the
calculations presented, the adiabatic index was assumed to be
close to unity (namely, $\gamma$ = 1.01), which approaches the
isothermic case [23]. This, in turn, allows us to treat the
$\gamma \sim 1$ model used as a system with energy losses. A
similar technique for taking into account radiative losses is
well known and has been applied in a number of studies (see, for
example, [16, 24]). This procedure clearly prevents accurate
allowance for radiative cooling of gas in the system, but,
nonetheless, brings the model used closer to a realistic case.

\subsection{Mathematical Model}

   We used a 3D system of gas dynamics equations, closed by the
ideal gas equation of state with adiabatic index $\gamma=1.01$,
to describe the gas flow. We used an integral form of this
system, since these equations allow, together with continuous
solutions, nonisoentropic solutions satisfying the
Rankin--Hugoniot conditions at discontinuities (shock waves and
contact discontinuities). The system has the form

$$
\frac{\partial}{\partial t} \int\limits_V \rho~dV
+ \int\limits_\Sigma \rho({\bf v}\cdot {\bf n}) d\Sigma = 0
$$
$$
\frac{\partial}{\partial t} \int\limits_V \rho{\bf v}~dV
+ \int\limits_\Sigma \rho{\bf v}({\bf v}\cdot {\bf n}) d\Sigma
+ \int\limits_\Sigma P~{\bf n} d\Sigma =
\int\limits_V {\rho} {\bf F}~dV
$$
$$
\frac{\partial}{\partial t} \int\limits_V \rho E~dV
+ \int\limits_\Sigma \rho h ({\bf v}\cdot{\bf n}) d\Sigma =
\int\limits_V {\rho} ({\bf F}\cdot{\bf v})~dV~~~.
$$

The specific external force {\bf F} includes the Coriolis force,
and also the attractive force of the two gravitational centers
and the centrifugal force (both described by the Roche
potential).  It has the form

$$
{\bf F} = -\grad\Phi+2[{\bf v}\times{\bf \Omega}]~~~,
$$

where the Roche potential can be written

$$
\Phi({\bf r}) = -\frac{G M_1}{|{\bf r} - {\bf r}_1|}
-\frac{G M_2}{|{\bf r} - {\bf r}_2|}
-\frac{1}{2}\Omega^2{\left({\bf r}-{\bf r}_c\right)}^2 ~~~~~.
$$

Here, ${\bf v}$ is the velocity, ${\bf \Omega} =
(0,0,\Omega)$, $\Omega = 2\pi/P_{orb}$, ${\bf r}_1$ and ${\bf
r}_2$ are the centers of the system components, and ${\bf r}_c$
is the center of mass of the system.

We adopted the following boundary conditions.

(1) The outflowing star fills its Roche lobe. Over the entire
surface of this component, the gas velocity is normal to the
surface. The gas speed is assumed to equal the local sound speed
$v = c_0 = 9 km/s$.

(2) The conditions of free matter outflow are set both on the
accretor and at the outer boundary.

(3) The density $\rho$ in the outflowing component's Roche lobe
was taken to be $\rho_0$. Note that the boundary value of the
density in the outflowing component does not affect the
solution, since the system of equations scales with $\rho$ and
$P$. We used an arbitrary value for $\rho_0$ in the
calculations.  However, when a specific system with a known mass
loss rate is considered, the real density values can be found
simply by adjusting the calculated density values according to
the scale implied by the ratio of the real and model densities
at the surface of the outflowing component.

  We took the gas in
the calculation region to initially have low density and be at
rest.  This gas is later displaced by the gas flowing outward
from the primary component.

\subsection{Numerical Model}

   A fundamental problem in the numerical realization of gas
dynamics models is the choice of a method for solution of the
corresponding system of equations. Currently, finite difference
methods recommend themselves most highly. Of the wide variety of
finite difference schemes, the most accurate are so-called
Godunov methods [25]. These are based on approximating the
initial gas dynamics equations in divergence form (corresponding
to the conservation laws), which makes it possible to obtain
correct values at discontinuities in the case of discontinuous
solutions. The important features of finite difference methods,
which, in many respects, determine the resulting adequacy of the
solutions, are the approximation viscosity and the monotonicity
of the scheme. Increasing the approximation order decreases the
numerical viscosity. It is possible to decrease the numerical
viscosity in a scheme with a given order using characteristic
properties of the initial system of gas dynamics equations,
which is hyperbolic.  This approach leads to so-called Total
Variation Diminishing (TVD) schemes, which have the property of
monotonicity (corresponding to the absence of nonphysical
oscillations of the solutions near discontinuities) [26].
Increasing the approximation order and introducing special flow
restrictions that make it possible to preserve the monotonicity
lead to high-order TVD methods with minimal approximation
viscosity [27]. In spite of their relative complexity,  such
schemes are currently the most promising for the numerical
realization of gas dynamics models.

   We used a high-order TVD scheme to numerically solve the
system of equations.  Special studies showed that this method
makes it possible to adequately reflect flow structures that
include shock waves and tangential discontinuities, without
leading to artificial oscillations or smearing of features in
the flow pattern. In addition, the method allows us to study
flows with large density gradients, which is especially
important when treating the influence of common envelope gas on
the flow structure.

   We solved the gas dynamics equations in a Cartesian
coordinate system specified as follows:

 - origin at the center of the outflowing star;

 - $X$ axis directed along the line connecting the star centers,
from the outflowing component to the accretor;

 - $Z$ axis directed along the rotation axis;

 - $Y$ axis defined so that the resulting system is right-handed.

  The calculation region is the parallelepiped
\linebreak $(-A..2A)\times
(A..A)\times(0..A)$; by virtue of the symmetry of the problem,
we carried out the calculations only for the upper half-space.
We used a non-uniform (denser in the accretor zone) difference
grid with $78\times 69\times 35$ nodes for the numerical
calculations.

   The system of equations was solved using a method for
transition to a steady state, from arbitrarily chosen initial
conditions to the entry of the solution to a stationary regime.
Note that the characteristic gas dynamics timescale for the
establishment of the flow for the system under consideration
(the ratio of the system's characteristic size $A$ to the
propagation velocity of perturbations $c_0$) is of the order of
three orbital periods. Thus, to ensure that the resulting
solution was stationary, we carried out the calculations over a
substantially longer time interval --- more than ten orbital
periods.  We checked the degree of stationarity of the solution
using both local and integrated characteristics of the flow.

\section{Calculation Results}

\figurenum{1a}
\begin{figure}
\plotone{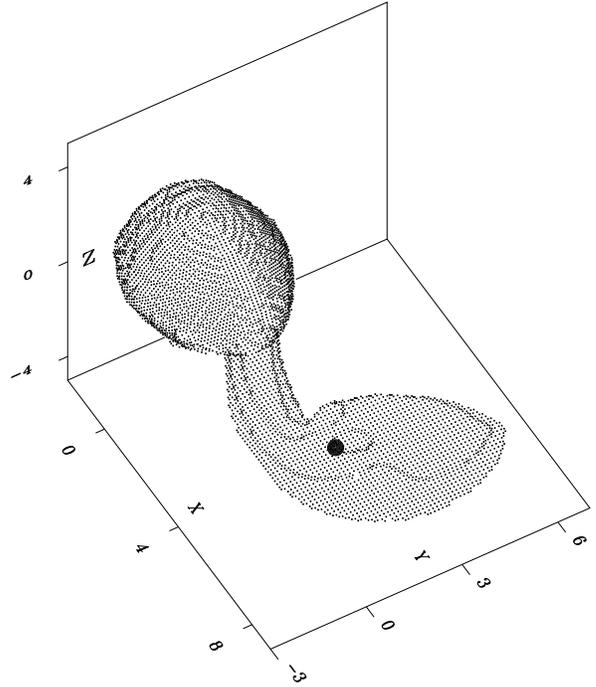}
\caption{3D view of density isosurface at the level $\rho =
0.01 \rho_0$.  $X$, $Y$, $Z$ are given in units of $R_\odot$.
The location of the accretor is marked by the dark circle.}
\end{figure}

   We will consider some characteristic features of the flow
structure in semidetached binary systems obtained using the 3D
gas dynamics model described in Section 2. The overall flow
pattern, which illustrates the morphology of gas flows in the
system, is shown in Fig.~1, which presents three-dimensional
images of density contours at the levels of 0.01 (Fig.~1a) and
0.05 (Fig.~1b) of the gas density at the point $L_1$. To better
show details of the calculated flow structure, Fig.~2 shows
cross sections of the density contours in Fig.~1a made by the
$XZ$ and $YZ$ planes passing through the accretor. The flow
pattern in Fig.~1 and Fig.~2 is time steady, and corresponds to the
steady flow regime achieved after a time interval exceeding ten
orbital periods from the starting time of the calculations.
These results reveal the following features of the calculated
flow structure.

\figurenum{1b}
\begin{figure}
\plotone{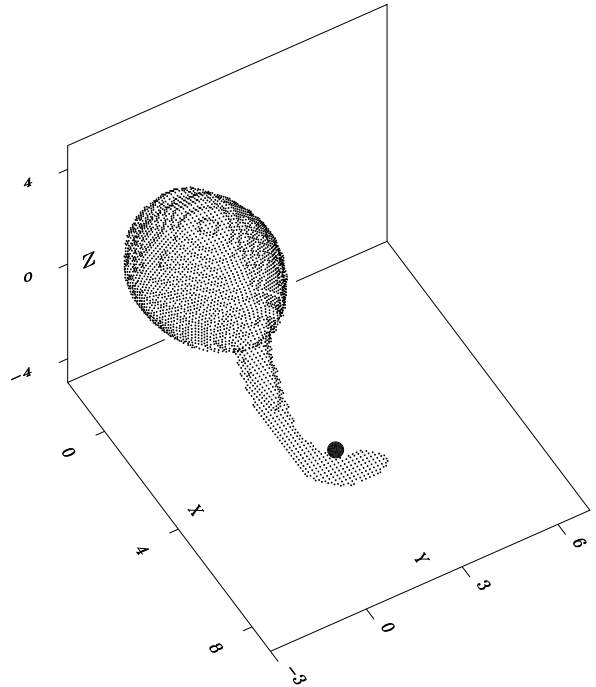}
\caption{3D view of density isosurface at the level $\rho =
0.05 \rho_0$.}
\end{figure}

\figurenum{2}
\begin{figure}
\plotone{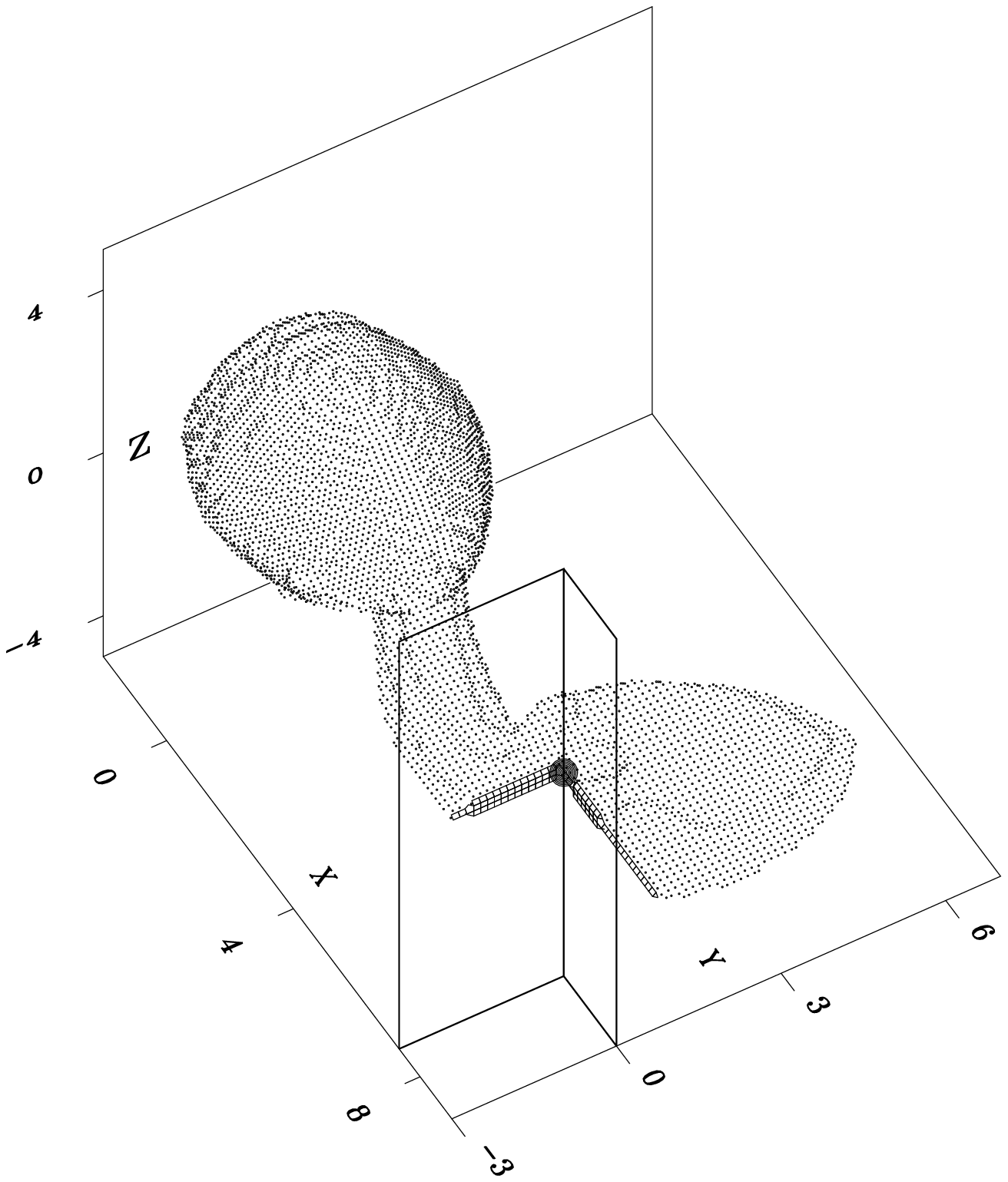}
\caption{The same density isosurface as on Fig.~1a,
sliced by the $XZ$ and $YZ$ half-planes passing through the accretor.}
\end{figure}

1) The matter of the stream can be conditionally divided into
three parts:  the first forms a quasi-elliptical accretion disk;
the second bends around the accretor and, remaining inside the
system, acts upon the overall flow pattern; and the third leaves
the system through the vicinity of Lagrange point $L_2$.

2) The interaction between the stream and the disk is not a
shock interaction.

3) The dimensions of the stream of matter flowing from the
vicinity of $L_1$ change as the stream propagates toward the
accretor: its height decreases and its width in the orbital
plane increases.

4) The height of the accretion disk that forms in the system is
much less than the height of the stream; this is confirmed by
the data presented in Fig.~3, which shows density contours up to
the value $\rho\sim 0.01\rho_0$ and velocity vector
in the $YZ$ plane, which forms a transverse cut through the disk
when it passes through the accretor.

\figurenum{3}
\begin{figure}
\plotone{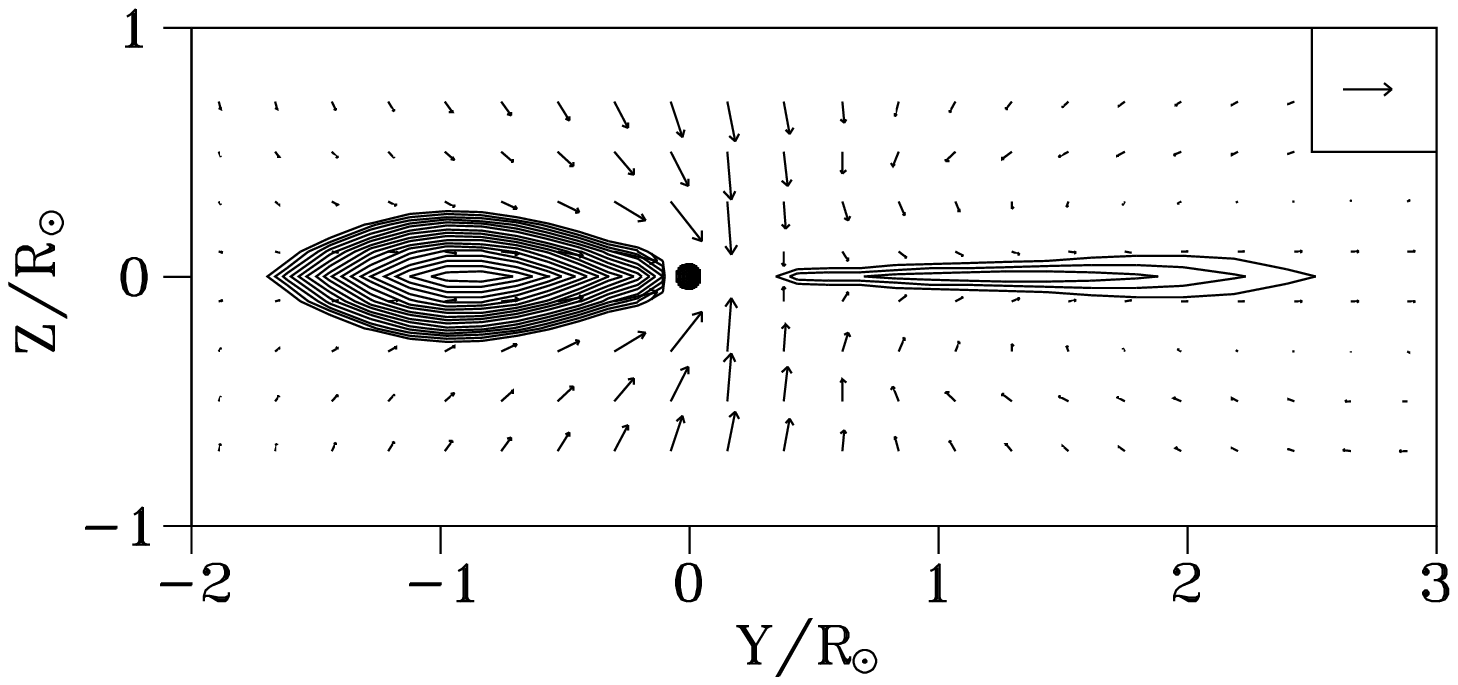}
\caption{Density isolines and velocity vectors in the plane $YZ$
passing through the accretor. The filled circle with the center
in the point $y=0$, $z=0$ marks the location of the accretor. Vector
in the upper right corner corresponds to the value of velocity of 1000
km/sec.}
\end{figure}

   We can analyze the structure of gas flows in more detail and
estimate the disk's linear size by considering the flow pattern
in the equatorial plane.  Figure 4 presents the density field
and velocity vector field in this plane inside an area with
dimensions from 2 to $10R_\odot$ along the $X$ axis and from -3
to $3R_\odot$ along the $Y$ axis. Four stream lines, marked with
the letters A, B, C, D and indicating the directions of flows in
the system, are also shown. Analysis of the results presented in
Fig.~4 confirms the above conclusion that some of the stream
material leaves the system through the vicinity of the Lagrange
point $L_2$ without getting close to the accretor (stream line
$A$). Some material falls directly onto the disk (stream line D)
and further, losing its angular momentum under the action of
viscosity, takes part in the accretion process. The quantitative
estimates obtained indicate that about 75% of the total amount
of gas injected into the system is accreted.

\figurenum{4}
\begin{figure}[t]
\plotone{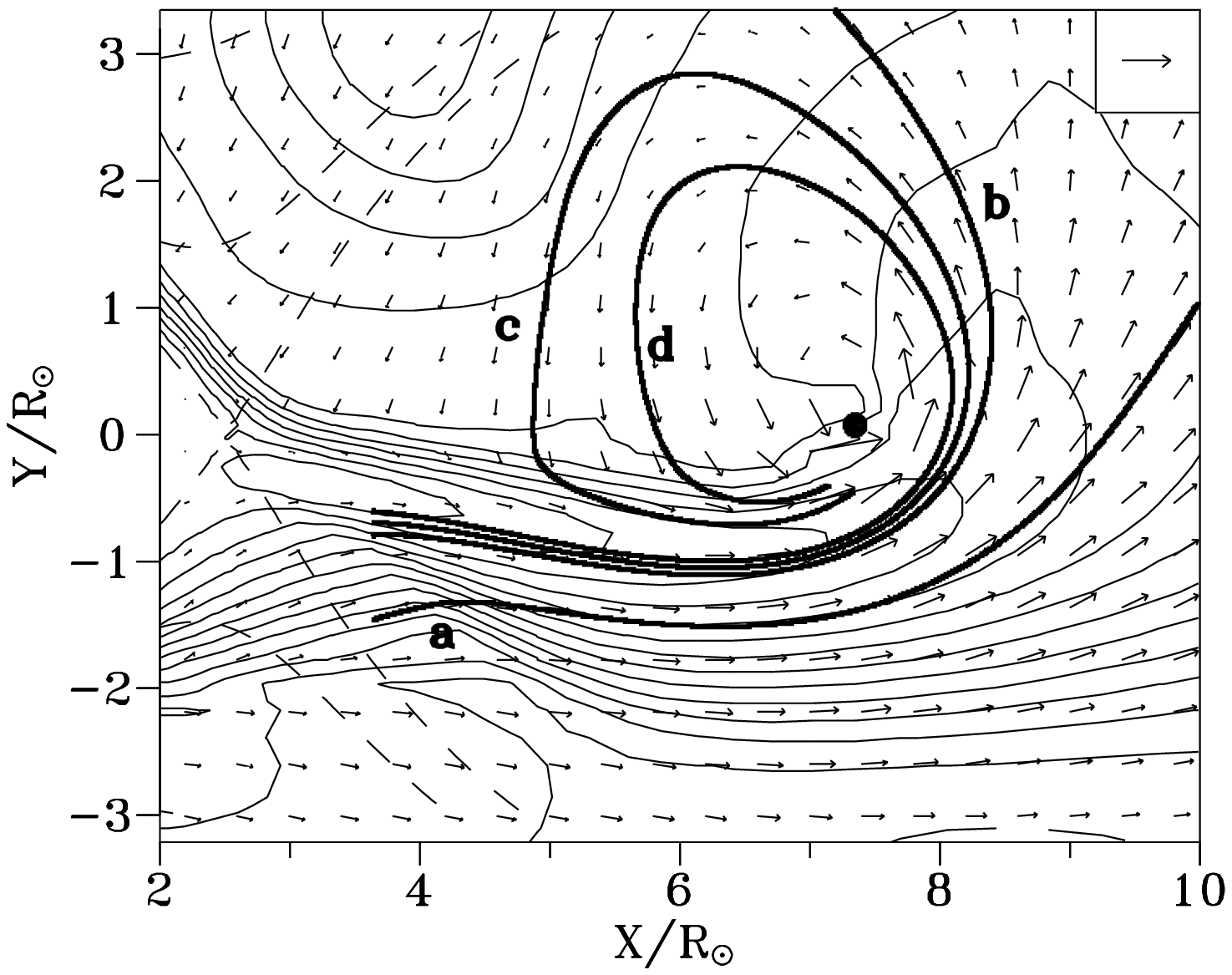}
\caption{
Density isolines and velocity vectors in the equatorial
plane of the system.
 The Roche equipotentials are shown with dashed lines. Four
stream lines marked with the letters A, B, C, and D are also shown. The
accretor position is marked with the dark circle. The vector in the upper
right corner corresponds to a velocity of 800 km/s.}
\end{figure}

   Of the most interest for the purposes of this paper is the
material that acts upon the flow pattern while remaining inside
the system (stream lines B and C in Fig.~4). We will call this
material the "common envelope" of the system. Note that a
significant part of the common envelope gas (see stream line B)
bends around the donor star and interacts with the matter
outflowing from its surface. The influence of this part of the
common envelope on the flow structure leads to substantial
changes in the mass exchange regime in the system. We will
provide a detailed consideration of this effect in a future
paper. Another part of the common envelope (see stream line C)
passes completely around the accretor and undergoes a shock
interaction with the edge of the stream facing into the orbital
motion. This type of interaction results in substantial changes
in the overall flow pattern, in particular, in the absence of a
"hot spot" in the disk and the formation of an extended shock
wave along the edge of the stream. We present a detailed
description of the impact of this part of the common envelope on the
gas flow morphology in the system below.

   To determine the linear dimensions of the disk, we must find
the limiting ("last") stream line along which matter flows
directly onto the disk. The preceding stream line (between lines
C and D), which directs matter around the accretor and then back
to interact with the stream, still passes outside the disk. The
matter along this line belongs to the common envelope and not to
the disk, although later (after interaction with the stream), it
can also be accreted. The limiting stream line in Fig.~4 is D,
and we can easily estimate the dimensions of the calculated
quasi-elliptical disk by examining this line: $3.2 \times 2.5
R_\odot$.  It follows from Fig.~3 that the disk thickness
increases with distance from the accretor, varying from $\sim
0.1$ to $\sim 0.2R_\odot$.  Note that to determine the disk
thickness more accurately, we must increase the number of grid
nodes, since the minimum calculated thickness of the disk was
limited by the size of a difference cell.

   The variations of the parameters for the gas flowing along
the stream lines presented in Fig.~4 show that the flow is
smooth along all lines belonging to the disk, up to the limiting
line D. The absence of discontinuities indicates a shockless
interaction between the stream and disk material, which, in
turn, implies the absence of a hot spot in the disk. The results
presented in Figs.~1, 2 and 4 provide an explanation for this
shockless morphology for the stream--disk system. Examination of
the flow pattern makes it clear that the stream deflected by the
action of the common envelope gas (stream line C in Fig.~4)
approaches the disk along a tangent, and, therefore, does not
cause any shock perturbation at the disk surface. At the same
time, the interaction of the stream with the common envelope
leads to the formation of an extended shock wave along the edge
of the stream facing into the orbital motion. We can estimate
the shock parameters, as well as the total amount of energy
released in the shock, using the calculation results.  One of
the characteristic features of the shock is its variable
intensity.  This is illustrated in Fig.~5, which presents the
distribution of the specific rate of energy release along the
shock in the equatorial plane $\delta E$, normalized to unity.
The vertical dashed lines in Fig.~5 show the edges of the
stream: the left line shows the point where the stream
originates (the point $L_1$), and the right line the place where
the stream ends, i.e., the place where the stream comes into
contact with the disk. We can easily seen in Fig.~5 that the
bulk of the energy in the system is released in the compact
shock region, adjacent to the accretion disk. This is extremely
important for interpretations of observational data, because the
compactness of the energy release region can explain the phase
relations of features in binary light curves in essentially the
same way as has been done previously assuming the presence of a
hot spot, in the framework of a fundamentally different model
for the matter flow in the system.

\figurenum{5}
\begin{figure}[t]
\plotone{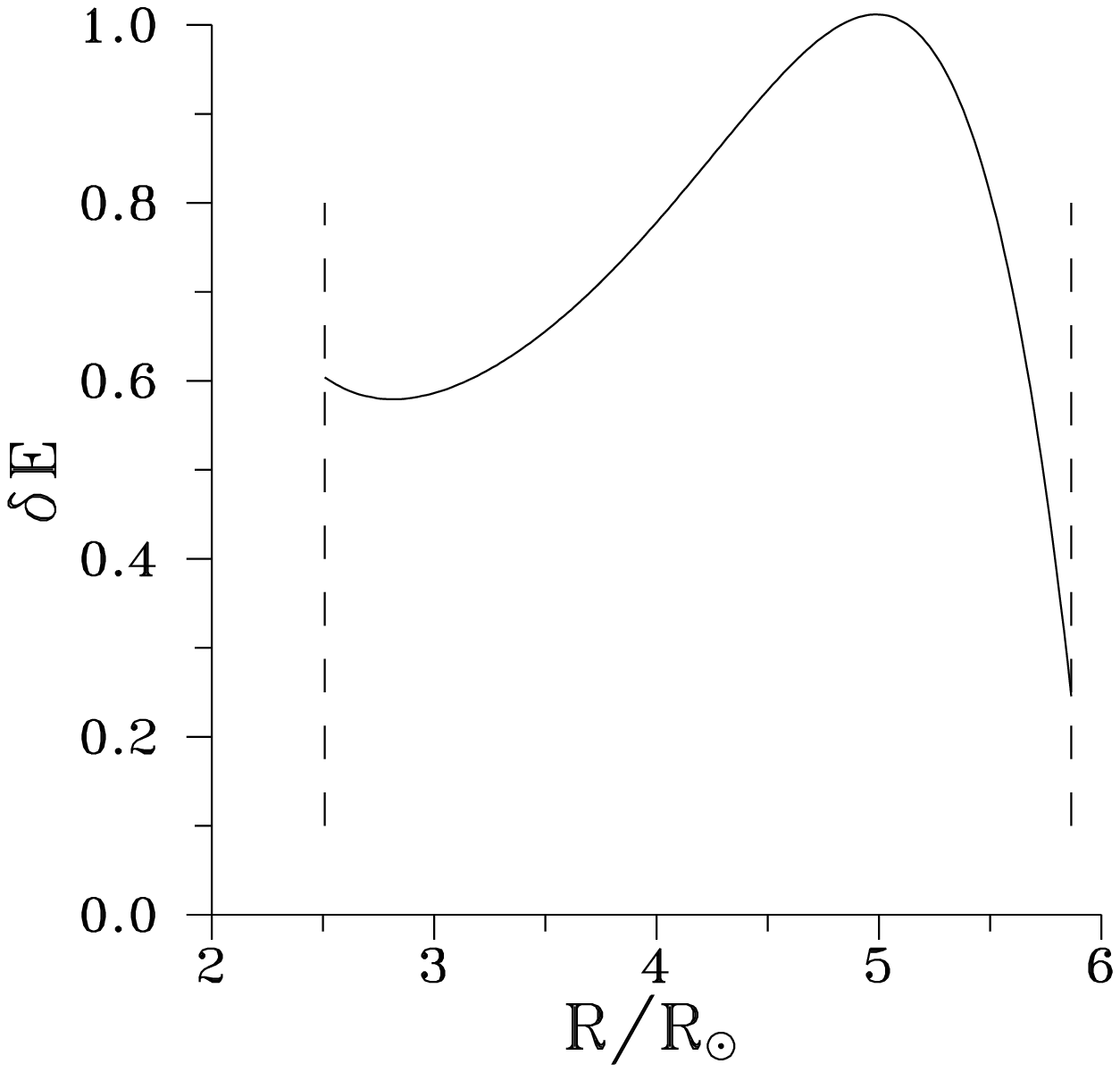}
\caption{
Distribution of the specific rate of energy release $\delta E$
along the shock wave in the equatorial plane, normalized to
unity. The dashed lines mark the shock boundaries.}
\end{figure}

   To verify that the shock forming along the edge of the stream
is an adequate substitution for the hypothetical hot spot in the
disk, let us consider the total energy release in both of these
formations.  In accordance with standard models, we assume the
hot spot to be generated at a distance $R_{out}$ from the
accretor.  In this case, the energy release resulting from the
stream and disk interaction is (see, e.g., [28]):

$$
\Delta E_{spot} = \frac{1}{4} \frac{G M_2 {\dot
M}}{R_{out}}~~~~~,
$$
where $\dot M$ is the mass loss rate of the donor star. For the
adopted values of the density $\rho_0$ and temperature (sound
speed $c_0$) at the star's surface, the value of $\dot M$ in
standard models is determined by the relation

$$
\dot M = \rho_0 c_0 S~~~,
$$
where $S$ is the cross-sectional area of the stream in the
vicinity of $L_1$ [3].  Using these formulas and determining
$R_{out}$ from the calculation results given above, assuming
the spot is generated at the point where the stream contacts the
disk, we estimated $\Delta E_{spot}$ for the gas density and
temperature at the surface of the donor star adopted in our
model. Comparison of the energy release rate in the shock
$\Delta E_{shock}$ with our $\Delta E_{spot}$ estimate shows
that $\Delta E_{shock}$ is twice $\Delta E_{spot}$. These
approximately equal energy release rates and the quite limited
size of the region where the bulk of the energy in the shock is
released ($\sim$ 60\% of the energy is released in the
one-third of the shock that is adjacent to the disk --- see
Fig.~5), suggests that the observational manifestations of this
shock will be roughly equivalent to those of a hot spot in the
disk.

Finally, we note that, in spite of the expected similarity of
the observational effects of the hypothetical hot spot and our
calculated shock wave, the calculated matter flow morphology in
semidetached binary systems is fundamentally different from the
standard picture, and this, in turn, calls for a reexamination
of a number of commonly accepted concepts. For instance, the
presence of a common envelope leads to significant changes in
the mass transfer rate in the system. In the case considered,
the common envelope causes matter to flow from a much (order of
magnitude) larger part of the surface of the donor star than is
supposed in standard models (this point will be considered more
thoroughly in a future paper). This means that for approximately
equal energy release rates in the different ("shock wave" and
"hot spot") models, the mass exchange rates will be
substantially different.  Correct concepts about the flow
morphology are especially important when interpreting
observations, since comparisons between model calculations and
observational data can provide information about the system.

\section{Conclusion}

   We have presented the results of 3D numerical simulations of
the matter flow pattern in a semidetached binary system similar
to the low mass X-ray binary X1822--371. Our results provide
evidence for the substantial influence of the rarefied common
envelope gas on the structure of gas streams in the system. The
common envelope gas interacts with the stream outflowing from
the vicinity of $L_1$ and deflects it, leading to a shockless
(tangential) interaction between the stream and the outer edge
of the accretion disk, and, consequently, to the absence of a
hot spot in the disk.  At the same time, the interaction of the
common envelope gas with the stream results in the formation of
an extended shock wave of varying intensity along the edge of
the stream. A preliminary analysis suggests that the
observational manifestations of this shock and the standard
hypothetical hot spot are similar, and, therefore, this shock
can be considered equivalent to a hot spot in the interpretation
of observations.

   Note that these results were obtained for a steady flow
regime.  In a non-stationary regime, when the flow morphology is
determined by external factors and is not self-consistent,
regions of shock interaction between the disk and the gas stream
are also possible. For example, if the disk is already formed
before the donor star fills its Roche lobe, a hot spot can be
created where the matter stream makes contact with the outer
edge of the disk after the mass exchange through the vicinity of
$L_1$ begins. Since a self-consistent solution without a hot
spot is expected after a steady-state flow regime is achieved,
it is important to determine the life time for the hot spot. It
is natural to adopt for its characteristic life time the time
required for the amount of matter injected into the system by
the stream to become comparable to the accretion disk mass,
since after complete mass exchange, the solution will become
self-consistent.  With mass exchange and accretion disk
parameters typical for the system under consideration [20], we
expect that the stationary flow regime will already be
established after a time of the order of 100 rotation periods.
This suggests that the probability of observing a hot spot is
extremely low, and all the observational manifestations that
have been associated with hypothetical hot spots are, in fact,
consequences of energy release in the shock on the stream edge.

In summary, we conclude that correctly taking into account the common
envelope in numerical simulations of the gas dynamics of mass transfer in
semidetached binary systems has revealed new features of the gas flow
structures, which significantly change our concepts of the morphology of
flows in these systems.

\acknowledgments

This work was supported by the Russian Foundation for Basic Research
(project code 96-02-16140).

\end{document}